\documentstyle[12pt]{article}

\def\vev#1{\langle#1\rangle}

\newcommand{\del}{\partial}

\def\Dsl{\,\raise.15ex\hbox{$/$}\mkern-13.5mu D} 
\def\phys{{\rm phys}}
\def\cont{{\rm cont}}
\def\latt{{\rm latt}}
\def\QCD{{\rm QCD}}
\def\MS{{\overline{\rm MS}}}

\def\Tr{\mathop{\rm Tr}\nolimits}
\def\GeV{\mathop{\rm GeV}\nolimits}
\def\MeV{\mathop{\rm MeV}\nolimits}
\def\fm{\mathop{\rm fm}\nolimits}

\begin{document}
\include{psfig}
\begin{titlepage}
 \null
 \begin{center}
 \makebox[\textwidth][r]{UW/PT-92-22}
 \makebox[\textwidth][r]{DOE/ER/40614-37}
 \vskip.5in
  {\Large
	RECENT PROGRESS IN LATTICE QCD}
  \par
 \vskip 2.0em
 {
  \begin{tabular}[t]{c}
\large    Stephen R. Sharpe\footnotemark \\[1.em]
        \em Department of Physics, FM-15\\
        \em University of Washington\\
        \em Seattle, Washington 98195
  \end{tabular}}
 \par \vskip .5in
 {\large\bf Abstract}
\quotation
I give a brief overview of the status of lattice QCD,
concentrating on topics relevant to phenomenology.
I discuss the calculation of the light quark spectrum,
the lattice prediction of $\alpha_\MS(M_Z)$,
and the calculation of $f_B$.
\endquotation
\end{center}
\vspace{.5in}
\begin{center}
   {\em Plenary talk given at the Meeting of the Division of Particles
	and Fields of the American Physical Society,
	Fermilab, November 10-14, 1992}
\end{center}
\vspace{1.in}

\centerline{PREPARED FOR THE U.S.~DEPARTMENT OF ENERGY}
\medskip
{\scriptsize

\noindent This report was prepared as an account of work
sponsored by the United States Government.  Neither the United States nor
the United States Department of Energy, nor any of their employees, nor
any of their contractors, subcontractors, or their employees, makes any
warranty, express or implied, or assumes any legal liability or responsibility
for the product or process disclosed, or represents that its use would not
infringe privately-owned rights.
By acceptance of this article, the publisher and/or recipient acknowledges
the U.S. Government's right to retain a nonexclusive, royalty-free license
in and to any copyright
covering this paper.

}

\footnotetext{email: sharpe@galileo.phys.washington.edu}
\vfill
\mbox{December 1992}
\end{titlepage}

\section{INTRODUCTION}

Lattice gauge theory has been somewhat out of the mainstream of
particle physics for the past decade.
It seems to me, however, that the field is now coming of age.
It has certainly grown rapidly: roughly 300 people attended the
recent {\em LATTICE '92} conference, compared to about 130 at
{\em LATTICE '86}.  That it has also matured is indicated
by the breadth of the subjects being studied with lattice methods.
These include the traditional---QCD spectrum and matrix elements; the
more statistical mechanical---rigorous theorems on finite size
scaling; the more abstract---random surfaces; and the exotic---finite
temperature baryon number violation.
More concrete evidence for maturation is that lattice results are
becoming useful to the rest of the particle physics community.
For example, in his summary of B-physics, David Cassel noted that the
bounds on the elements of the CKM-matrix which follow from $B-\bar B$ mixing
depend on the values of $f_B$ and $B_B$, and that lattice results
are now used as one of the estimates of these numbers.
But perhaps the most important piece of evidence is that lattice studies
are beginning to produce results with no unknown systematic errors.
The best example is the calculation of the full QCD running coupling constant
$\alpha_\MS$ by the Fermilab group \cite{fnalalpha}.
As I discuss below, this number can be directly compared to experiment.

In summary, I would say that
lattice studies are beginning to make themselves useful.
The ``beginning'' in this claim is important---there
is a very long way to go before we can calculate, say, the $K\to\pi\pi$
amplitudes from first principles.
Thus this talk does not consist of results for a long list of matrix elements.
Instead, I discuss a few topics in some detail.
Because of lack of time,
I concentrate entirely on results from lattice QCD (LQCD)
which are relevant to phenomenology.

When thinking about the progress that has been made,
it is useful to keep in mind the questions that one ultimately
wishes to answer using LQCD:
\begin{enumerate}
\item
Does QCD give rise to the observed spectrum of hadrons?
Do these particles have the observed properties
(charge radii, decay widths, etc)?
\item
Does the same theory also describe the perturbative, high-energy
jet physics?
\item
What are the values of the hadronic matrix elements
($f_B$, $B_K$, etc.)
which are needed to determine the poorly known elements of the CKM matrix?
\item
What happens to QCD at finite temperature, and at finite density?
\item
What are the properties of
``exotic'' states in the spectrum, e.g. glueballs?
\item
What is the physics behind confinement and chiral symmetry breaking?
\end{enumerate}
The last question is the hardest, and, while it is important to keep
thinking about it, there have not been any breakthroughs.
The other questions are being addressed using numerical simulations,
complemented by a variety of analytic calculations.
Most recent progress has concerned the first four questions,
and I will discuss only the first three.
For additional details see the reviews of  Mackenzie (heavy quarks),
Petersson (finite temperature), Sachrajda (weak matrix elements)
and Ukawa (spectrum) at {\em LATTICE '92} \cite{lat92}.

To set the stage I begin with a brief summary of LQCD, eschewing as many
details as possible.
The action of QCD is the sum of a gauge term,
\begin{equation}
\label{gluonaction}
S_g = {6\over g^2} \int_x \frac{1}{12}\sum_{\mu\nu}
\Tr(G_{\mu\nu} G_{\mu\nu}) \ ,
\end{equation}
with $G_{\mu\nu}$ the gluon field strength, and a quark part
\[
S_q = \int_x \sum_q \bar q (\gamma_\mu D_\mu + m_q) q \ , \ \
D_\mu = \del_\mu + i A_\mu .
\]
These expressions are valid in Euclidean space,
where numerical lattice calculations are almost always done.
QCD is put on a hypercubic lattice by placing the quarks on the sites,
and gauge fields on the links which join these sites,
in such a way that gauge invariance is maintained.
The derivative in $D_\mu$ can be discretized in many ways,
leading to different types of lattice fermions.
The most common choices are  Wilson and staggered fermions.
The coupling constant $g$ has been absorbed in the gauge fields,
and appears only as an overall factor in $S_g$.
It is conventional (see Eq. \ref{gluonaction})
to use the combination $\beta=6/g^2$ to specify $g$.
In present simulations $\beta\sim6$, so that $g^2\sim 1$.
The lattice spacing $a$ is determined implicitly by the choice of $g^2$,
as discussed below.

The prototypical quantity of interest is the two-point correlator, e.g.
\begin{eqnarray}
\nonumber
C(t) &=& \left\langle \sum_{\vec x}
[\bar b \gamma_0 \gamma_5 d(t,\vec x)]\
[\bar d \gamma_0 \gamma_5 b(0)]\right\rangle \\
\label{twopteqn}
&=& {\rm sign}(t)\ f_B^2 m_B e^{-m_B|t|}
\left( 1 + O(e^{-(m_{B'}-m_B)|t|}) \right) \ .
\end{eqnarray}
At large Euclidean time, $t$, one picks out the lightest state
(here the $B$ meson). The contribution of excited states
(beginning with the $B'$) is suppressed exponentially.
Thus one can just read off the mass, $m_B$,
while the amplitude of the exponential gives the decay constant
($f_B$) up to kinematical factors.

\begin{figure}[tb]
\centerline{\psfig{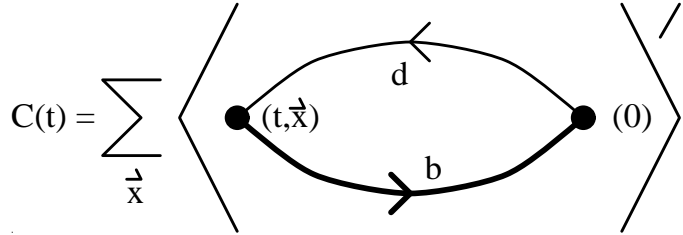}}
\caption[figeq3]{Diagrammatic representation of Eq. \ref{newtwopteqn}}
\label{figeq3}
\end{figure}

The expectation value in Eq. \ref{twopteqn} indicates a
functional integral over quarks, antiquarks and gluons
weighted by $\exp(-S_g-S_q)$.
Doing the quark and antiquark integrals, one obtains
\begin{equation}
\label{newtwopteqn}
C(t) = - \int [dA]\ e^{-S_g}\
[\Pi_q {\rm det}(\Dsl+m_q)]\ \sum_{\vec x}
\Tr[\gamma_0\gamma_5 G_d(t,\vec x;0)\gamma_0\gamma_5 G_b(0;t,\vec x)]
\ .
\end{equation}
Here $[dA]$ is shorthand for the functional integral over lattice
gauge fields, and the $G$ are quark propagators
$ G_q = (\Dsl + m_q)^{-1} $.
The integral is normalized to give unity in
the absence of the trace term.
I have shown Eq. \ref{newtwopteqn} diagrammatically in Fig. \ref{figeq3},
where the lines are quark propagators,
and the primed expectation value includes the determinant.

To make the calculations numerically tractable, they must be done
in a finite volume of $N_s^3\times N_t$ points.
The functional integral is then of large but finite dimension,
and can be done by Monte-Carlo methods. Similarly, the propagators
are obtained by inverting a finite matrix.
Available computer speed and memory limits the number of sites
in this four dimensional world.
In fact, it is speed that limits present simulations:
the bottleneck is the inclusion of ${\rm det}(\Dsl + m_q)$ in the measure.
To make progress, many calculations use the ``quenched'' approximation
in which the determinant in Eq. \ref{newtwopteqn} is set equal to unity.
This is equivalent to dropping all internal quark loops,
so that there are only valence quarks.
I discuss the consequences of this drastic approximation below.

It is important to realize that numerical LQCD calculations
which do not use the quenched approximation
become exact when the lattice spacing vanishes, the volume goes to
infinity, and the statistical errors vanish.
They are thus non-perturbative calculations from first principles,
which have errors that can be systematically reduced.
They are not ``model'' calculations.
On the other hand, calculations in the quenched approximation are more
similar to those in a model. Certain physics is being left out,
and one does not know {\em a priori} how large an error is being made.
As I will show, present calculations suggest that the error is small,
at least for a range of quantities.

In the following three sections I discuss the spectrum, the calculation
of $\alpha_\MS$, and that of $f_B$. I will focus mainly on
quenched results since these are more extensive.
I will only make some brief comments on results for ``full QCD'',
i.e. QCD including the determinant in the measure.

\section{SPECTRUM}

I begin by discussing the status of calculations of the spectrum
of light hadrons. I concentrate on $m_\pi$, $m_\rho$ and $m_N$
($N$ refers to the nucleon), since these have the smallest errors.
To give an idea of the rate of progress, I will
compare with results from March 1990 \cite{sspascos}.
The two and a half year interval since then
is long enough to allow substantial changes.
For example, computer power has increased as roughly
$CPU\propto e^{\rm year}$,
so that, for a four dimensional theory such as QCD,
the linear dimensions could have increased by a factor of
$\sim e^{2.5/4}=1.8$ in this period.
In fact, the extra time has only partly been used in this way.

\begin{figure}[t]
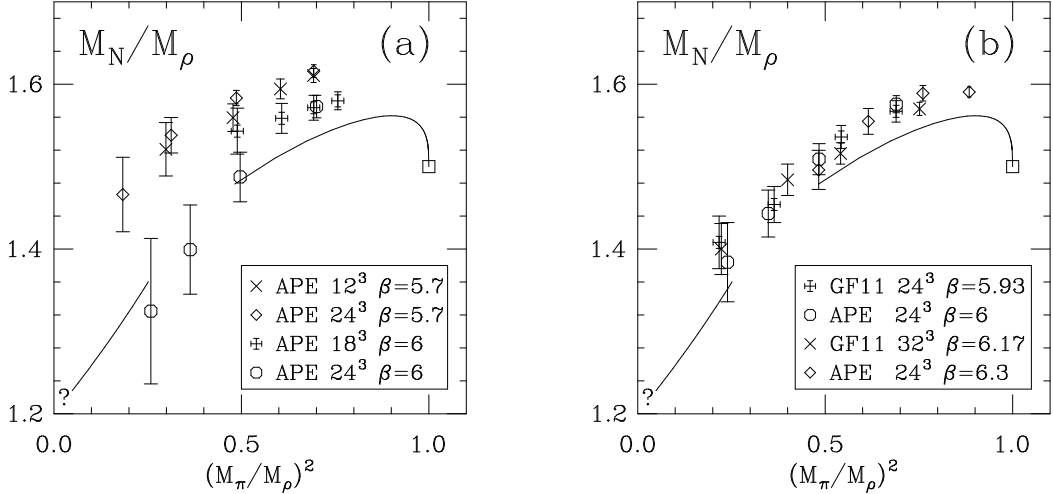

\vspace{2.5truein}
\includegraphics{fig1a.ps}
\includegraphics{fig1b.ps}
\caption[figapew]{APE plots for quenched Wilson fermions:
(a) March 1990 (b) November 1992}
\label{figapew}
\end{figure}

To display the data I follow the ``APE'' group and
plot $m_N/m_\rho$ versus $(m_\pi/m_\rho)^2$ (see Figs. \ref{figapew}).
To understand the significance of these plots, recall the following.
In a lattice calculation, we can dial the values of the quark masses.
Ignoring for the moment the strange quark, and assuming degenerate
up and down quarks, we then have a single light quark mass, $m_l$,
at our disposal. Each value of $m_l$ corresponds to a possible
theory, each with different values for dimensionless mass ratios such as
$m_\pi^2/m_\rho^2$, $m_\rho/m_N$, $f_\pi/m_\rho$, $m_\Delta/m_N$, etc.
We would like to fix $m_l$ using one of these ratios,
and then predict the others.
In practice, it is technically very difficult to do a simulation
with small enough $m_l$, and so we must extrapolate.
The APE plot is one way of displaying how well this extrapolation
agrees with the experimental masses.
As $m_l$ varies from $0\to\infty$, $m_\pi^2/m_\rho^2$ varies
monotonically from $0\to1$.
Thus the theory maps out a curve in this plot,
which we know must pass through the infinite mass point
($m_\pi=m_\rho=2 m_N/3$; shown by a square in the plots).
The issue is whether this curve passes through the experimental point,
indicated by a ``?'' on the plots.

The solid lines if Figs. \ref{figapew} are the predictions of
two phenomenological models.
The curve for light quark masses uses
chiral perturbation theory, with certain assumptions,
and is constrained to go through the physical point.
The curve at heavier masses is based on a potential model.
For more discussion see Ref. \cite{sspascos}.

What do we expect for the lattice results?
There will be corrections to physical quantities which vanish
in the continuum limit as powers of $a$
(up to logarithms, which I will ignore), e.g.
\begin{equation}
\label{eqcontlimit}
(m_N/m_\rho)_\latt = (m_N/m_\rho)_\cont [1 + a \Lambda + O(a^2)] \ .
\end{equation}
Thus, for finite $a$, the lattice curve should not
pass through the experimental point.  Similarly, if the physical size
of the lattice ($L=N_s a$) becomes too small the masses will be
shifted from their infinite volume values.
Thus, in addition to extrapolating in the quark mass,
one must attempt to extrapolate both $a\to 0$ and $L\to\infty$.

\begin{table}[tb]
\caption[tabapew]{\tenrm\baselineskip=12pt
Parameters of lattices used to produce data shown in Fig. \ref{figapew}}
\begin{center}
\renewcommand{\arraystretch}{1.2}
\begin{tabular}{cclccccc}
Year 	& Ref.	& $\beta$ & $a(\fm)$
			 &$\pi/a(\GeV)$	& $N_s$& $L(\fm)$	& Lattices \\
\hline
1990&\cite{ape90}& 5.7	& 1/6  & 3.8 	& 12,24	& 2,4		& 294,50 \\
    &\cite{ape90}& 6	& 1/10 & 6.3	& 18,24 & 1.8,2.4	& 104,33 \\
\hline
1992&\cite{weingarten}
    		 & 5.93 & 1/9  & 5.7	& 24	& 2.7		& 217 \\
    &\cite{ape6n}& 6	& 1/10 & 6.3	& 24	& 2.4		& 78 \\
    &\cite{weingarten}
    		 & 6.17 & 1/12 & 7.5	& 32 	& 2.7		& 219 \\
    &\cite{ape63}& 6.3	& 1/16 & 10	& 24	& 1.5		& 128 \\
\hline
\end{tabular}
\end{center}
\label{tabapew}
\end{table}

What has happened in the last two years is that these extrapolations
have become more reliable. I illustrate this with results using Wilson
fermions in the quenched approximation
(details of the data set are given in Table \ref{tabapew}).
Figure \ref{figapew}a shows the state-of-the-art in March 1990.
The upper two sets of points are from a large lattice spacing
($a\approx 1/6\fm$) with two different lattice sizes ($L\approx2$ and $4\fm$).
The results agree within errors, and I concluded that we knew the
infinite volume curve for $a=1/6\fm$ with a precision of about $5\%$.
This curve appeared {\em not} to pass through the physical point.
The lower two sets of points are from a smaller lattice spacing
($a\approx1/10\fm$). They suggested a downward shift in the data
with decreasing lattice spacing, but it was difficult to draw
definite conclusions given the large errors.

The present results are shown in Fig. \ref{figapew}b.
To avoid clutter, I show only data with $a\le1/9\fm$.%
\footnote{%
The new results at $a=1/6\fm$ \cite{weingarten} confirm, with reduced errors,
those in Fig. \ref{figapew}a.}
These are all consistent with a single curve lying below that at $a=1/6\fm$
and close to the phenomenological predictions.
The improvements in the results have come from using
more lattices to approximate the functional integral,
which reduces the statistical errors (see Table \ref{tabapew}).
Furthermore, the decrease in $a$ has been compensated by an increase in $N_s$,
so that (with the exception of the APE results at $\beta=6.3$)
the physical extent of the lattices exceeds $L=2\fm$.
The results at $\beta=5.7$ imply that this is large enough
to get within a few percent of the infinite volume results.

The GF11 group have used their results at $a\approx1/6,1/9,1/12\fm$
to do an extrapolation both to physical quark masses and to $a=0$
\cite{weingarten}.
They find that, for a range of quantities, the results
are consistent with their experimental values within $\sim 5\%$ errors.
For example, $m_N/m_\rho=1.284{+0.071 \atop -0.065}$ (cf $1.222$ expt.)
and $m_\Delta/m_\rho=1.627{+0.051\atop - 0.092}$ (cf $1.604$ expt.).
It should be recalled that a few years ago the quenched results
for $m_N/m_\rho$ were thought to be larger than the experimental value,
while $m_\Delta-m_N$ was smaller.
What we have learned is that {\em both} the errors introduced by working in
finite volume and at finite lattice spacing shift the curve in the
APE plot upwards, so that one can easily be misled by results on small
lattices.

We seem, then, to know the spectrum of light hadrons
in the quenched approximation fairly well,
and it looks a lot like the observed spectrum.
There is not, however, complete agreement on the numerical results.
The QCDPAX collaboration,
working at lattice spacings comparable to those in Fig. \ref{figapew}b,
finds significantly larger values for $m_N/m_\rho$ at smaller
$m_\pi^2/m_\rho^2$ \cite{qcdpax}.
They suggest that the disagreement may be due to a contamination
from excited states in the correlators (see Eq. \ref{twopteqn}).
This disagreement will get cleared up in the next year or so.
What is needed is operators which couple more strongly to the ground state,
and less to excited states.
There has been considerable improvement in such operators,
for systems involving a heavy quark \cite{eichtenlat92},
but less so for light quark hadrons.

Let me assume that the results of Fig. \ref{figapew}b are correct,
so that the quenched spectrum does agree with the experiment to within
$\sim 5\%$. Does this imply that we
can trust quenched calculations of other quantities to this accuracy?
I do not think so.
{\em A priori}, I would not have expected the quenched approximation
to work so well, because it leaves out so much important physics.
For example, the quenched rho cannot decay into two pions,
unlike the physical rho, which
might lead to a 10\% underestimate of its mass \cite{geiger}.
Also, the pion cloud around the light hadrons is much different in the
quenched approximation than in full QCD,
which should shift $m_\rho$ and $m_N$ in different ways.
While it is possible that these effects largely cancel for the spectrum,
I see no reason for them to do so in other quantities.
This argument can be made more concrete for the charge radii \cite{leinweber}.
Futhermore, as I mention below,
there are reasons to think that the approach to the chiral limit
in the quenched approximation is singular.

My final remark on the quenched spectrum concerns the possibility of
improving the approach to the continuum limit \cite{symanzik}.
The gauge action is accurate to $O(a^2)$, but the Wilson fermion action
has $O(a)$ corrections, as in Eq. \ref{eqcontlimit}.
It is possible to systematically ``improve'' the fermion action so
that the corrections are successively reduced by powers of $g^2/4\pi$.
The idea is that by using a more complicated action one can work
at a larger lattice spacing.
The first results of this program are encouraging.
Ref. \cite{apeimproved} finds that an improved action shifts the results
at $a=1/6\fm$ downwards so as to agree with those at $a\approx1/10\fm$,
and that both agree with
the ``unimproved'' results at $a\le1/9\fm$ shown in Fig. \ref{figapew}b.
Ref. \cite{ukqcd} finds that the improved action makes no difference at
a smaller lattice spacing, $a\approx 1/14\fm$.
All this suggests that it may be possible use lattice spacings
as large as $1/6\fm$ with an improved Wilson fermion action.
Further evidence for this comes from lattice studies
of charmonium \cite{aida,ukqcdhyperfine}.
With staggered fermions, on the other hand, the prospects are less rosy.
Although the corrections are of $O(a^2)$, and thus parametrically smaller
than with Wilson fermions, they are in fact large at $a=1/6\fm$ \cite{sslat91}.

Ultimately, we must repeat the calculation using full QCD.
Much of the increase in computer time in the last few years has gone
into simulations which include quark loops.
Nevertheless,  it is too early to discuss physical predictions,
since it is not yet possible to reliably extrapolate $m_q\to 0$ or $a\to0$.
The limit $L\to\infty$ is, however, well understood and some
interesting results have been obtained.
The Kyoto-Tsukuba group finds good fits to the form
$m = m_{\infty}(1+c/L^3)$,
and argue that this can be understood as due to the
``squeezing'' of the hadrons in the finite box \cite{fukufinitevol}.%
\footnote{This is not the asymptotic form:
for large enough $L$ the power law becomes an exponential \cite{luscher}.}
The MILC collaboration finds that at $L=2.7\fm$ the finite volume effects
are smaller than $2\%$ \cite{milc}.
Both these results are consistent with the finite volume effects
observed in the quenched approximation \cite{weingarten}.

\section{$\alpha_\MS$ FROM THE LATTICE}

It is straightforward, in principle, to extract $\alpha_\MS$
from lattice calculations:
\begin{enumerate}
\item
Pick a value of $\beta=6/g^2$.
\item
Calculate a physically measurable quantity, e.g. $m_\rho$ or $f_\pi$.
\item
Compare lattice and physical values and extract $a$ using, e.g.\footnote{%
The values for $a$ quoted above were obtained in this way, using
a variety of physical quantities.}
\begin{equation}
\label{eqlattphys}
(m_\rho)_\latt =  (m_\rho)_\phys \times a \ (1 + O(a))\ .
\end{equation}
The $O(a)$ terms are not included when extracting $a$.
\item
Convert from the lattice bare coupling constant $g^2(a)$ to
$\alpha_\MS(q\!=\!\pi/a)$ using perturbation theory.
The result can then be evolved to other scales, e.g. $M_Z$, using the
renormalization group equation.
\item
Repeat for a variety of values of $\beta$, and extrapolate to
the continuum limit $a=0$. In this way the $O(a)$ corrections
in Eq. \ref{eqlattphys} are removed.
\end{enumerate}
If this program were to be carried through,
then the lattice result for $\alpha_\MS$ would allow
absolute predictions of jet cross sections, $R(e^+e^-)$, etc.
(modulo the effects of hadronization).
If these predictions were successful, it
would demonstrate that QCD could simultaneously explain
widely disparate phenomena occurring over a large range of mass scales.
While such success has not yet been achieved, there has been considerable
progress in the last year or so.
I will explain the two major problems,
and the extent to which they have been resolved.
I then discuss the present results.

\subsection{Reliability of Perturbation Theory}

The fourth step in the program requires that
perturbation theory be valid at the scale of the lattice cut-off,
which is roughly $\pi/a$ in momentum space.
On present lattices this ranges from $5-12 \GeV$
(the values are given in Table \ref{tabapew}).
It turns out that these values are not large enough for perturbation
theory in the bare coupling constant to be accurate,
because there are large higher order corrections.
These are exemplified by the relation needed in step 4 above
\cite{hasenfratz}
\begin{equation}
\label{eqalphaMS}
{1\over\alpha_\MS(\pi/a)} = {1\over\alpha_\latt(a)} - 3.880 + O(\alpha) \ ;
\ \ \ (\alpha=g^2/4\pi) \ .
\end{equation}
Since $\alpha_\latt\approx 1/13$, the first order correction is large,
and higher order terms are likely to be important.

This problem has been understood by Lepage and Mackenzie
\cite{lepagemackenzie}.
The large corrections arise from fluctuations of the lattice gauge fields,
and in particular from ``tadpole'' diagrams which are present on the lattice
but absent in the continuum.
The solution is to express perturbative results in terms of a
``continuum-like'' coupling constant,
e.g. $\alpha_{\rm MOM}$ or $\alpha_\MS$, with the scale $q\approx\pi/a$.
When this is done the higher order coefficients are considerably reduced.
This is similar to what happens in the continuum when
one shifts from the MS to the $\MS$ scheme.

Having re-expressed all perturbative expressions in terms of, say,
$\alpha_\MS$, there remains the problem of finding the value of this
coupling in terms of $\alpha_\latt$,
since Eq. \ref{eqalphaMS} is not reliable.
One has to use a non-perturbative definition of coupling
constant which automatically sums up the large contributions,
and which is related to $\alpha_\MS$ in a reliable way.
One choice is \cite{lepagemackenzie}
\begin{equation}
\label{newalphaMS}
\alpha_P = - {3 \ln \vev{\Tr U_P} \over 4 \pi} \ ;\ \
{1\over\alpha_\MS(\pi/a)} = {1\over\alpha_P} - 0.5 + O(\alpha) \ ,
\end{equation}
where $U_P$ is the product of gauge links around an elementary square.
One first determines $\alpha_P$ from the numerical value of $\Tr U_P$,
and then converts this to $\alpha_\MS$,
which is then used in perturbative expressions.
Lepage and Mackenzie find that the resulting
numerical predictions of lattice perturbation theory work well
for quantities (such as small Wilson loops)
which are dominated by short-distance perturbative contributions.
This is true for lattice spacings as large as $1/9\fm$, and perhaps $1/6\fm$.
Thus the determination of $\alpha_\MS$ using Eq. \ref{newalphaMS} is
reliable, with errors probably no larger than a few percent.

\subsection{Errors Introduced by the Quenched Approximation}

The dominant source of uncertainty in present calculations of $\alpha_\MS$
is the use of the quenched approximation.
The problem is the lack of a
``physical'' quenched theory to use in the comparison of step 3 above.
This shows up in two ways. First, the value of $a$ depends on the
quantity chosen in the comparison: using $m_\rho$ gives one value,
$f_\pi$ another.\footnote{%
Actually, since light hadron mass ratios are well reproduced by the
quenched approximation, the variation of $a$ is small for such quantities.
This need not be true in general.}
Second, the coupling constant that one obtains is
for a theory with zero flavors, $\alpha_\MS^{(0)}$,
and must somehow be related to the physical coupling constant $\alpha_\MS$.
However, such a relationship involves non-perturbative physics,
and so can only be determined by a calculation using full QCD!
The best that we can hope for at present is a good estimate of the
relationship between the couplings.

Recently, the FNAL group have made such an estimate for the coupling
determined using the $1P-1S$ splitting in charmonium \cite{fnalalpha}.
The crucial simplifying feature is that
charmonium is described reasonably well by a potential model,
so one need only estimate the effects of quark loops
on the potential itself. In outline, this is done as follows.
Matching the lattice and continuum $1P-1S$ splittings makes the
quenched and full QCD potentials similar at separations $R\sim 0.5\fm$.
The potentials will, however, differ at smaller separations.
We understand this difference at small enough $R$,
where the Coulomb term dominates, i.e. $V\propto-\alpha_\MS(1/R)/R$.
In the quenched approximation $\alpha$ varies more rapidly with $R$
because of the absence of fermion loops,
which means that the quenched potential is steeper.
Assuming that this is true all the way out to $0.5 \fm$,
where the potentials match,
implies that the quenched potential must lie below that for
full QCD at short distances.
It follows that $\alpha_\MS > \alpha_\MS^{(0)}$ at short distances.
A quantitative estimate is given in Ref. \cite{fnalalpha}.

Unfortunately, there is no such simple way of estimating
the effects of the quenched
approximation on the values of $\alpha_\MS^{(0)}$ extracted from the
properties of light quark hadrons.

\subsection{Results}

Various physical quantities have been used to calculate $\alpha_\MS^{(0)}$,
and I collect the most accurate results in Table \ref{alphatab}.
I also give the ``experimental'' number obtained from an average of
various perturbative QCD fits to data \cite{pdt}.
I quote the coupling at the scale $M_Z$,
to allow comparison with Fig. 1 of the
1992 Review of Particle Properties (RPP) \cite{pdt}.
The second row gives the results of the FNAL group,
including the correction for quenching.
The remaining results use the ``string tension'', $\sigma$.
This is the coefficient of the linear term in the heavy quark-antiquark
potential: $V(R)\to \sigma R$ for $R\to\infty$.
All groups use the ``improved'' perturbation theory explained above.
I have not shown results from light hadron masses as they are less accurate.

\begin{table}
\caption[alphatab]{\tenrm\baselineskip=12pt
Results for $\alpha_\MS$ in quenched and full QCD. Errors are statistical.}
\renewcommand{\arraystretch}{1.2}
\begin{center}
\begin{tabular}{ccll}
Quantity 	& Ref. 		& $\alpha_\MS^{(0)}(M_Z)$
					& $\alpha_\MS(M_Z)$ 	\\[.2em]
\hline
``Experiment''	& \cite{pdt}	&	    & 0.1134(35) \\
$M_{1P}-M_{1S}$ & \cite{fnalalpha}
				& 0.0790(9) & 0.105(4)	\\
String tension	& \cite{bali} 	& 0.0796(3) & 	\\
String tension 	& \cite{ukqcdas}& 0.0801(9) & 	\\ \hline
\end{tabular}
\end{center}
\label{alphatab}
\end{table}

The results for $\alpha_\MS^{(0)}$ obtained using $\sigma$
have the smallest statistical errors.
Indeed, it is a triumph of LQCD that the
linear term in the quenched potential is very well established,
for it is this term that causes confinement.
The difficulty is that $\sigma$ is not a physical quantity.
The ``physical'' value, $\sqrt\sigma=0.44\GeV$, is extracted
from the potential needed to fit the $\bar c c$ and $\bar b b$ spectra.
These systems do not, however, probe the region of the potential where
the linear term dominates. Furthermore, at large $R$ the full QCD potential
flattens out due to quark-pair creation, while the quenched potential
continues its linear rise.
Thus the extracted value of $\sigma$ is somewhat uncertain,
and it is difficult to relate $\alpha_\MS^{(0)}$ obtained
using $\sigma$ to the full QCD coupling.
Nevertheless, it seems to me possible to make a rough estimate,
and it would certainly be interesting to try.

The lattice prediction for $\alpha_\MS$
is slightly less than $2\sigma$ below the RPP average.
In his summary talk, Keith Ellis quoted an updated average,
$\alpha_\MS(M_Z) = 0.119(4)$,
which is almost $3\sigma$ higher than the lattice result.
I think that this near agreement is a success of LQCD,
and that it is too early to make anything out of the small discrepancy.
For one thing, the dominant error in the lattice result
is the uncertainty in the conversion from $\alpha_\MS^{(0)}$ to $\alpha_\MS$,
and this might be underestimated.
It is important to realize, however, that this uncertainty will be gradually
reduced as simulations of full QCD improve.

I end this section with a general comment on the calculations.
For the method to work, both non-perturbative and
perturbative physics must be included {\em in a single lattice simulation}.
At long distances the quarks are confined in hadrons, while at the
scale of the lattice spacing their interactions
must be described by perturbation theory.
On a finite lattice, these two requirements pull in opposite directions.
For example, if perturbation theory requires $a<1/9\fm$,
while finite volume effects require $L>2.7\fm$ (as may be true for light
hadrons), the lattice must have $N_s\ge30$, and so present lattices are
barely large enough.
This is another reason for using charmonium to determine $\alpha_\MS$.
The $\bar c c$ states are smaller than light hadrons,
and it turns out that $N_s=24$ is large enough to reduce the
finite volume errors below 1\% \cite{fnalalpha}.

Nevertheless, it would be nice to extend the range of scales so as to
provide a detailed test of the dependence on $a$.
To accomplish this with present resources requires
the use of multiple lattices having a range of sizes.
L\"uscher {\em et al.} have proposed such a program,
and done the calculation for SU(2) pure gauge theory \cite{luscheralpha}.

\section{The B-meson decay constant $f_B$}

One of the most important numbers that LQCD can provide phenomenologists
is $f_B$. Analyses of the constraints due to $\bar K-K$ and $\bar B-B$ mixing
find (for $m_t\ge 140 \GeV$) two types of solutions for CP-violation in
the CKM-matrix (see, e.g. Ref. \cite{lusignoli}).
These are distinguished by whether $f_B$ is
``small'' ($100-150 \MeV$) or ``large'' ($160-340 \MeV$).
In the former case, CP-violation in $B\to K_S J/\psi$ is small;
in the latter it is much larger.
We would like the lattice to resolve this ambiguity.

The calculation of $f_B$ is also interesting as a testing ground for
the ``heavy quark effective theory'' (HQET) \cite{isgurwise}.
The B-meson consists of heavy $b$ quark, and a light quark ($u$ or $d$).
Imagine that we were free to vary the heavy quark mass, $m_Q$.
The mass of the pseudoscalar meson
(which I call $m_P$ to distiguish it from the physical $B$ meson mass, $m_B$),
and its decay constant $f_P$, would depend upon $m_Q$.
As $m_Q\to\infty$ one can show that \cite{eichten}
\begin{equation}
\label{hqeteqn}
\phi_P = f_P \sqrt{m_P} \left[{\alpha(m_P)\over \alpha(m_B)}\right]^{2/\beta_0}
= \phi_\infty \left(1 + {A\over m_P} + {B\over m_P^2} + \dots \right)
\ .
\end{equation}
Here $\beta_0$ is the first term in the QCD $\beta$-function,
and $A$ and $B$ are constants except for a weak logarithmic
dependence on $m_P$.
The issue for HQET is the size of the $1/m_P$ corrections
at the $B$ and $D$ masses, for these might be indicative of the size of the
corrections in other applications of HQET, e.g. $B\to D$ transitions.

It is important to realize that, while the $m_Q\to\infty$ limit
simplifies the kinematics of the heavy quark,
it does not simplify the dynamics of the light quark.
In particular, the quenched approximation is no better in the
heavy quark limit, nor is the dynamics more perturbative.
Thus one needs a LQCD calculation here just as much as for light quark
hadrons.

\begin{figure}[t]
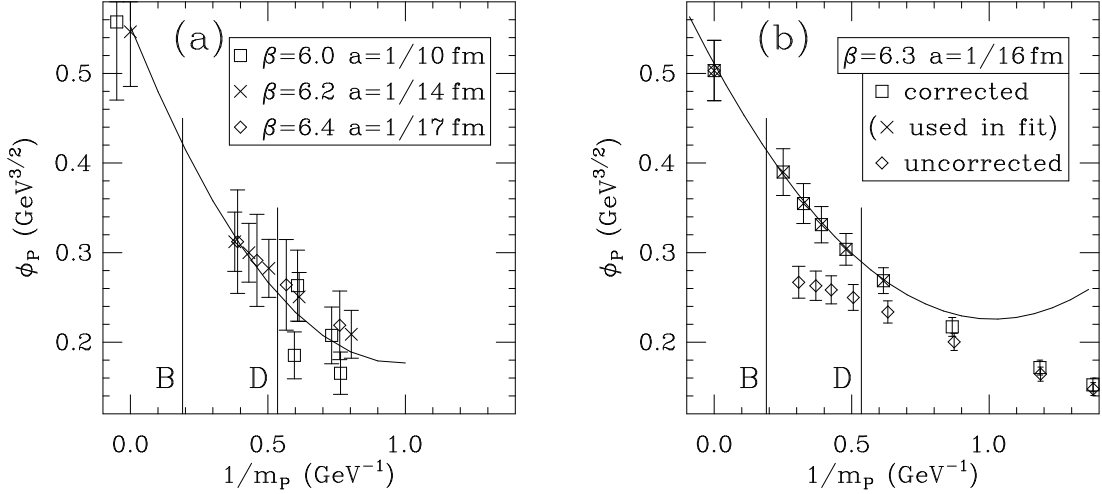

\vspace{2.5truein}
\includegraphics{fig2a.ps}
\includegraphics{fig2b.ps}
\caption[fBfig]{Results for $\phi_P=m_P\sqrt{m_P}$:
(a) from Ref. \cite{abada}, (b) from Ref. \cite{labrenz92}.}
\label{fBfig}
\end{figure}

What we would like to do is map out the curve of $\phi_P$
versus $1/m_P$, and read off the values of $f_B$ and $f_D$.
Examples of present results are shown in Figs. \ref{fBfig}, and
in Table \ref{fBtab} I give the present numerical values
for $f_B$ and $f_D$.
I also quote $f_B({\rm static})=\phi_\infty/\sqrt{m_B}$,
which is the value of $f_B$
ignoring the $1/m_P$ corrections in Eq. \ref{hqeteqn}.
Unfortunately, the situation is less straightforward than the figures
imply, and I will spend the remainder of this section explaining
and evaluating these results.

There are three major causes of uncertainty in $f_B$.
The first concerns the overall scale.
To extract a result in physical units we need to know the lattice spacing.
As discussed above, the value we obtain depends on the physical quantity
that we use, particularly at finite lattice spacing.
$f_B$ is more sensitive to the uncertainty in $a$
than are light hadron masses,
since one is calculating $a^{3/2}\phi_P$ rather than $a m$.
To illustrate this sensitivity, I include in Table \ref{fBtab}
results from Refs. \cite{eichtenlat92,dina}
using two different determination of $a$.
For the other results, the uncertainty is about $15\%$.
Ultimately, to remove this uncertainty one must repeat the
calculation with full QCD.

The second problem concerns the isolation of the lightest state.
In principle, calculating $\phi_P$ is straightforward.
One simply studies the long time behavior of the two point function
of the local axial current, Eq. \ref{twopteqn}, and reads off $f_P \sqrt{m_P}$.
In practice, to obtain a signal it is necessary to use extended operators,
which couple more strongly to the lightest state than the local current.
This is particularly important at large $m_Q$.
With a less than optimal operator there are likely to be systematic
errors introduced by contamination from excited states.
Indeed, there are disagreements between results using different operators.
This is illustrated in Table \ref{fBtab} by the variation in
$f_B({\rm static})$.
It seems to me, however, that the operators of Ref. \cite{thacker}
are close to optimal, and that the
discrepancies will go away as other groups optimize their operators.

\begin{table}
\caption[fBtab]{\tenrm\baselineskip=12pt
 Results for decays constants in the quenched approximation.
The normalization is such that $f_\pi=132\MeV$.
Only statistical errors are shown;
systematic errors are discussed in the text.}
\renewcommand{\arraystretch}{1.2}
\begin{center}
\begin{tabular}{ccclll}
Ref.	&$\beta$& Scale from	& $f_B({\rm static})(\MeV)$
					& $f_B(\MeV)$& $f_D(\MeV)$\\[.1em]
\hline
\cite{abada}
	&6.0-6.4& $f_\pi,m_\rho$& 310(25)	& 205(40)	& 210(15) \\
\cite{ukqcdfB}
	&6.2	& $f_\pi$	&		& 183(10)	& 198(5) \\
\cite{labrenz92}
	&6.3	& $f_\pi$	& 230(15)	& 187(10)	& 208(9) \\
\cite{eichtenlat92}
	&5.9 & $M(1P)\!-\!M(1S)$& 319(11) 	& 	& \\
	&	& $\sigma$	& 265(10) 	& 	& \\
\cite{dina}
      &5.74-6.26& $\sigma$	& 230(22)	&	& \\
	&	& $m_\rho$	& 256(78)	&	& \\
\hline
\end{tabular}
\end{center}
\label{fBtab}
\end{table}

The third and most difficult problem concerns putting very heavy
quarks on the lattice. As the quark's mass increases, the ratio
of its Compton wavelength to the lattice spacing, $1/(m_Q a)$, decreases.
For $m_Q a > 1$, its propagation through the lattice will be severely
affected by lattice artifacts.
There are, however, no such difficulties for an infinitely massive quark,
for such a quark remains at rest both in the continuum and on the lattice
\cite{eichten}.
Thus it seems that we are forced to interpolate between
$m_Q \sim 1/a $ and $m_Q=\infty$.

To illustrate the problems that this introduces, consider the situation
about two years ago. The smallest lattice spacing was
$a=1/10\fm$, so that $m_c a\approx 0.75$ and $m_b\approx 2.5$.
Typical results are those represented by squares in Fig. \ref{fBfig}a.
The quark mass has been restricted to satisfy $m_Q a\le 0.7$ to avoid
large artifacts (an arbitrary but reasonable choice for an upper bound),
and thus lie to the right of the $D$ meson line.
Clearly it is difficult to convincingly interpolate using Eq. \ref{hqeteqn}:
the variation in $\phi_P$ is so large that one cannot truncate the
$1/m_P$ expansion. Nevertheless, if one assumes that the curvature is
not too large, the fact that we know $\phi_\infty$  does
allow a rough estimate of $f_B$.

There has been considerable progress in the last two years.
The lattice spacing has been reduced to $1/17\fm$,
which allows one to use heavier quarks while keeping $m_Q a < 1$.
This is illustrated by the remaining points
in Fig. \ref{fBfig}a, all of which have $m_Q a\le 0.7$.
These points can now be fit in a more reliable way to the asymptotic
form of Eq. \ref{hqeteqn}. The curve shows such a fit, the results
from which are given in Table \ref{fBtab}.
Other groups, however, find results in apparent disagreement.
For example. the ``uncorrected'' points in Fig. \ref{fBfig}b
should agree with those in \ref{fBfig}a, but instead are lower.
This disagreement is, I suspect, mainly due to inadequate
isolation of the lightest $B$ meson by one or both groups.

Another development has been the use an improved fermion action
\cite{ukqcdfB}.
Since this has smaller $O(a)$ corrections, one should be able
to work at larger values of $m_Q a$.
I give the results in Table \ref{fBtab}.

Despite these improvements, it would be much better if we could work at
any value of $m_Q a$, and simply map out the entire curve.
There is a dispute about whether this is possible,
and I will attempt to give a summary of the arguments.

I begin by noting that the errors in $\phi_P$ do
not keep growing as $m_Q a$ increases.
This is because, if $m_Q\gg \Lambda_{\rm QCD}$, the quark is non-relativistic.
Its dynamics can then be expanded in powers of $1/m_Q$ \cite{nrqcd}
\begin{equation}
\label{nrqcdeqn}
{\cal L} = \psi^{\dag} \left[i D_0 + {D_i^2\over 2m_1} +
c(g^2){\sigma_i B_i \over 2 m_2} + O(1/m_Q^2) \right] \psi \ ,
\end{equation}
where $\psi$ is a two component field, and $c$ is a perturbative coefficient.
In the continuum, $m_1=m_2=m_Q$. The discrete rotational symmetries
are sufficient to ensure that a heavy lattice quark will be described by
the same Lagrangian, except that neither $m_1$ nor $m_2$
are equal to the bare lattice quark mass $m_Q$.
Nevertheless, although the lattice heavy quarks have the wrong dynamics,
this should only introduce errors of $O(\Lambda_\QCD/m_Q)$.
Thus, if $a$ is small enough that $m_Q a < 1$ when the quark becomes
non-relativistic, the errors will be small for all $m_Q a$.

Strictly speaking, to bring the lattice Lagrangian into the form of
Eq. \ref{nrqcdeqn},
one must perform a wavefuction renormalization on the heavy field,
which changes the normalization of the $\bar b u$ axial current.
This has been calculated in perturbation theory
keeping only the ``tadpole'' diagrams which are
thought to be dominant \cite{kronfeld}.
The result is illustrated in Fig. \ref{fBfig}b:
upon renormalization the ``uncorrected'' points are shifted upwards
into those labeled ``corrected''.
With this modification, the curve for finite $m_Q a$ is guaranteed to
pass through the $\phi_\infty$. Without it, the curve will
bend over and eventually pass through the origin.
In fact, the uncorrected points in Fig. \ref{fBfig}b
do appear to be flattening out at the smallest values of $1/m_P$,
although those of Fig. \ref{fBfig}a do not.

A second correction has also been applied in Fig. \ref{fBfig}b.
It is possible to partly account for the difference between
$m_1$ and $m_Q$ by an appropriate shift to smaller $1/m_Q$.
The size of this shift has been calculated keeping
``tadpole'' diagrams \cite{kronfeld}.
This correction ensures that the data approach $\phi_\infty$ with
a slope which is linear in $1/m_P$.
This slope will not, however, be correct since
the $1/m_2$ term in Eq. \ref{nrqcdeqn} has the wrong normalization.
Nevertheless, in contrast to the ``uncorrected'' points,
the shifted points fit well to the form of Eq. \ref{hqeteqn}.
The fit is shown by the curve, and
the resulting values for decay constants are given in Table \ref{fBtab}.

The controversial issue is how to further modify the calculation
so as to obtain the correct $1/m_P$ terms.
Kronfeld, Lepage and Mackenzie have suggested a program for removing the error
\cite{kronfeld}.
The idea is to add new terms to the lattice action,
and to choose the parameters so that
one obtains the correct Lagrangian at $O(1/m_Q)$.
This is essentially a complicated way of putting non-relativistic QCD
on the lattice \cite{nrqcd}.
The difficult question in both cases is how to fix the parameters.

Kronfeld {\em et al} propose that this can be done using perturbation theory.
Maiani {\em et al.} argue, however, that non-perturbative
contributions may be important \cite{maiani}.
Normally, such contributions are suppressed by powers of $a$,
but here they are enhanced by factors of $1/a$ due to linear divergences.
(A clear explanation of this effect is given in Ref. \cite{guido}.)
The result is that, with parameters fixed using perturbation theory,
there will be errors in $\phi_P$ which are of $O(\Lambda_\QCD/m_Q)$,
i.e. of the same size as the terms one is trying to calculate.
The point of disagreement is whether these non-perturbative terms are
large or small.
If they are large, the only solution would be to
fix parameters using non-perturbative normalization conditions.
In general this would reduce predictive power.

Clearly more work is needed to resolve this dispute.
One way to do this is to push the tests of perturbation theory
\cite{lepagemackenzie}
to the level at which non-perturbative terms show up.

It is fortunate, however, that this uncertainty has only a small
effect on $f_B$ and $f_D$.
This is shown by the good agreement in Table \ref{fBtab}
despite the different methods being used.
The results favor the ``large'' solution for $f_B$.
The only way this could change is if the systematic error
due to the quenched approximation turns out to be large.

We can also estimate the size of the $O(1/M)$ corrections
to the heavy quark limit. Taking the data of Ref. \cite{labrenz92}
as an example, they are $\sim 15\%$ for $f_B$,
and $\sim 45\%$ for $f_D$.
These numbers increase further if one uses
the larger values of $\phi_\infty$ found by Ref. \cite{eichtenlat92}.

\section{OUTLOOK}

There are many interesting developments that I have not have had time or
space to cover.
One which impinges on much of the work discussed above
concerns the accuracy of the quenched approximation \cite{eta'}.
In this approximation there are nine pseudo-Goldstone bosons,
rather than the eight of QCD, the extra one being the $\eta'$.
This means that quenched particles have an $\eta'$ cloud,
not present in full QCD.
It turns out that this gives rise to singularities in the chiral limit,
much as pion loops give singularities in the
charge radii of pions and nucleons \cite{leinweber}.
For example, the quark condensate diverges.
The implications of these divergences are not yet clear.
The most optimistic view is that the effects of $\eta'$ loops
are small as long as we work above a certain quark mass.
This is supported by the absence of numerical evidence to date
for the divergences.

Unfortunately, the quenched approximation is likely to be with us for a
number of years. At present, the fastest computers simulating LQCD are
running at $1-10$ GFlops. Even a TeraFlop machine,
such as that proposed by the TeraFlop collaboration \cite{teraflop},
will focus on quenched lattices.
These will have $N_s\approx100$,
and should give definitive quenched results for a reasonable number of
interesting quantities.
The major problem with simulations of full QCD
is that the CPU time scales as $m_\pi^{-10.5}$ with present algorithms.
Clearly, it is crucial that effort go into improving these algorithms.

\section*{ACKNOWLEDGEMENTS}

I thank Don Weingarten Jim Labrenz and Akira Ukawa for providing me with data,
and Estia Eichten, Aida El-Khadra, Rajan Gupta, Brian Hill, Andreas Kronfeld,
Jim Labrenz and Paul Mackenzie for discussions.
This work was supported by the DOE under contract DE-FG09-91ER40614,
and by an Alfred P. Sloan Fellowship.

%
\def\etal{{\it et al.}}
\def\PRL#1#2#3{{\it Phys. Rev. Lett.} {\bf #1}, #3 (#2) }
\def\PRD#1#2#3{{\it Phys. Rev.} {\bf D#1}, #3 (#2)}
\def\PLB#1#2#3{{\it Phys. Lett.} {\bf #1B} (#2) #3}
\def\NPB#1#2#3{{\it Nucl. Phys.} {\bf B#1} (#2) #3}
\def\NPBPS#1#2#3{{Nucl. Phys.} {\bf B ({Proc. Suppl.}){#1}} (#2) #3}
\def\brookhaven{Proc. {\it ``Lattice Gauge Theory '86''},
             Brookhaven, USA, Sept. 1986, NATO Series B: Physics Vol. 159}
\def\ringberg#1{Proceedings of the Ringberg Workshop,
	{\it ``Hadronic Matrix Elements and Weak Decays''},
        Ringberg, Germany, 1988, edited by A. Buras \etal,
	\NPBPS{7A}{1989}{#1}}
\def\seillac#1{
  Proceedings of the International Symposium on Lattice Field Theory,
  {\em ``Field theory on the Lattice''},
  Seillac, France, 1987, edited by A. Billoire \etal,
  \NPBPS{4}{1988}{#1}}
\def\fermilab#1{
  Proceedings of the International Symposium on Lattice Field Theory,
  {\em ``LATTICE 88''}
  Fermilab, USA, 1988, edited by A.S. Kronfeld and P.B. Mackenzie,
  \NPBPS{9}{1989}{#1}}
\def\capri#1{
  Proceedings of the International Symposium on Lattice Field Theory,
  {\em ``LATTICE 89''},
  Capri, Italy, 1989, edited by N. Cabibbo \etal,
  \NPBPS{17}{1990}{#1}}
\def\talla#1{
  Proceedings of the International Symposium on Lattice Field Theory,
  {\em ``LATTICE 90''},
  Tallahassee, Florida, USA, 1990, edited by U. M. Heller \etal,
  \NPBPS{20}{1991}{#1}}
\def\tsukuba#1{
  Proceedings of the International Symposium on Lattice Field Theory,
  {\em ``LATTICE 91''},
  Tsukuba, Japan, 1991, edited by M. Fukugita \etal,
  \NPBPS{26}{1992}{#1}}
\def\amsterdam{
  Talk at the International Symposium on Lattice Field Theory,
  {\em ``LATTICE 92''}, Amsterdam, The Netherlands, 1992,
  to be published in Nucl. Phys. {\bf B ({Proc. Suppl.}) } }

\end{document}